\documentclass[12pt]{iopart}

\usepackage{graphicx}   
\begin{document}
\title[Poynting flux in the neighbourhood of a point charge and the radiative power losses]
{Poynting flux in the neighbourhood of a point charge in arbitrary motion and the radiative power losses} 
\author{Ashok K. Singal}
\address{Astronomy and Astrophysics Division, Physical Research Laboratory,
Navrangpura, Ahmedabad - 380 009, India }
\ead{asingal@prl.res.in}
\vspace{10pt}
\begin{indented}
\item[]October 2015
\end{indented}
\begin{abstract}
We examine the electromagnetic fields in the neighbourhood of a ``point charge'' in arbitrary motion and thereby determine the Poynting flux 
across a spherical surface of vanishingly small radius surrounding the charge. We show that the radiative power losses from a point charge turn 
out to be proportional to the scalar product of the instantaneous velocity and the first time-derivative of the acceleration of the charge. 
This may seem to be in discordance with the familiar Larmor's formula where the instantaneous power radiated from a charge is proportional to 
the square of acceleration. However, it seems that the root cause of the discrepancy actually lies in the Larmor's formula which is derived 
using the acceleration fields but without a due consideration for the Poynting flux associated with the velocity-dependent self-fields 
``co-moving'' with the charge. Further, while deriving Larmor's formula one equates the Poynting flux through a surface at some later 
time to the radiation loss by the enclosed charge at the retarded time. Poynting's theorem, on the other hand, relates the outgoing radiation 
flux from a closed surface to the rate of energy decrease within the enclosed volume, all calculated {\em for the same given instant} only. 
Here we explicitly show the absence of any Poynting flux in the neighbourhood of an instantly stationary point charge, implying no radiative 
losses from such a charge, which is in complete conformity with energy conservation. We further show how Larmor's formula is still able to 
serve our purpose in vast majority of cases. It is further shown that Larmor's formula in general violates momentum conservation and in the 
case of synchrotron radiation leads to a potentially wrong conclusion about the pitch angle changes of the radiating charges, and that only 
the radiation reaction formula yields a correct result, consistent with the special relativity.

\end{abstract}
\section{Introduction}
In classical electrodynamics there is an old unsolved enigma -- what exactly constitutes the instantaneous rate of 
radiative power loss from an arbitrarily moving point charge -- 
is it proportional to the square of the acceleration (Larmor's formula) or is it proportional to the instant velocity 
multiplied with the rate of change of the acceleration, as inferred from the radiation reaction formula? 
This puzzle, a paradox, has now survived without a proper, universally acceptable, solution for more than a century.   
The conventional wisdom is that something may be lacking in the radiation losses derived from the radiation-reaction formula, 
which is thought to be perhaps not as rigorously derived as Larmor's formula, the latter derived 
by calculating the Poynting flux across a spherical surface of a large enough radius centred on the charge \cite{1,2,3,25}. 
The Poynting flux appears to be independent of the radius of the sphere if only the acceleration-dependent term is used, 
and one usually assigns this to be the power lost by the charge into radiation. 
We demonstrate here that a mathematical subtlety has been missed in the application of Poynting's theorem to derive Larmor's formula and that a 
proper mathematical procedure shows that the radiation losses can be considered to take place only when there is a 
change in the acceleration of the charge. For this we derive the power loss formula from the Poynting flux in the neighbourhood of the 
charge. We show that the radiative power is proportional to the scalar product of the instantaneous velocity and the 
first time-derivative of the acceleration of the charge. This expression for power loss was hitherto obtained in the literature 
only from a detailed derivation of the self-force of an accelerated charge sphere of a vanishingly small radius \cite{16,7,24,20}.
But here we have derived the power loss directly from the Poynting flow for a ``point charge''. 

That Larmor's formula could lead 
to wrong conclusions can be seen by examining the case in the instantaneous rest-frame of an accelerated charge. 
Such a charge has no velocity at that instant and hence no kinetic energy 
that could be lost into radiation. Even if some external agency, assumedly imparting acceleration to the charge, 
were considered to be supplying the energy going into the radiation, it could not have provided the power necessary for radiation 
in this case since work done by this external agency will also be zero as the system has a zero velocity at that instant.
However, according to Larmor's formula, the radiated
power is directly proportional to {\em square of the instantaneous value of the acceleration} of the charged particle, 
even for an instantly stationary charge. We  
explicitly calculate the Poynting flux passing through a spherical surface of vanishingly  
small dimensions surrounding the charge, in its instantaneous rest-frame, and from that we show that the Poynting flux in 
{\em the rest frame} is zero, indicating thereby the absence of radiative losses, in conformity with energy conservation. 
This in turn removes the need for the acceleration-dependent internal-energy term, introduced in the literature on an ad hoc 
basis \cite{7,41,56}, to comply with law of energy conservation.
We shall further demonstrate that in the instantaneous rest-frame of a {\em uniformly accelerated charge}, Poynting flux is zero 
not only in its immediate neighbourhood, but at all radial distances from the charge. 
This in turn implies an absence of radiation for a charge supported in 
a gravitational field which is in conformity with the strong principle of equivalence.
Finally we would show that how Larmor's 
formula could lead to a wrong inference about the dynamics of a charge radiating by a synchrotron process, in particular about 
its pitch angle changes as the charge loses energy by radiation, and where only the radiation reaction formula leads to a correct result.

\section{Larmor's formula for radiation from an accelerated charge}
The electromagnetic field (${\bf E},{\bf B}$)  of an arbitrarily moving charge $e$ is given by \cite{1,28},
\begin{equation}
\label{eq:1a}
{\bf E}=\left[\frac{e({\bf n}-{\bf v}/c)} {\gamma ^{2}r^{2}(1-{\bf n}\cdot{\bf v}/c)^{3}} + 
\frac{e{\bf n}\times\{({\bf n}-{\bf v}/c)\times \dot{\bf v}\}}{rc^2\:(1-{\bf n}\cdot {\bf v}/c)^{3}}\right]_{ret},
\end{equation}
\begin{equation}
\label{eq:1b}
{\bf B}={\bf n} \times {\bf E}\:,
\end{equation}
where the quantities in square brackets are to be evaluated 
at the retarded time. More specifically, ${\bf v}$, $\dot{\bf v}$, and 
$\gamma=1/\surd(1-(v/c)^{2})$ represent respectively the velocity, acceleration and the Lorentz
factor of the charge at the retarded time, while ${\bf r}=r{\bf n}$ is the radial vector from the retarded 
position of the charge to the field point where electromagnetic fields are being evaluated.

To calculate the electromagnetic power (radiation!) passing through a spherical surface $\Sigma$ around the charge, 
we make use of the radial component of the Poynting vector \cite{1}, 
\begin{equation}
\label{eq:8a}
{\bf n} \cdot {\cal S}= \frac{c}{4\pi}{\bf n} \cdot ({\bf E}\times {\bf B})
=\frac{c}{4\pi}({\bf n}\times {\bf E})\cdot {\bf B}=\frac{c}{4\pi}({\bf n}\times {\bf E})^2.
\end{equation}

Using only the acceleration fields (second term on the right hand side in Eq.~(\ref{eq:1a})), which are transverse in nature 
(perpendicular to ${\bf n}$) and assuming a non-relativistic motion, one gets for the Poynting flux, 
\begin{equation}
\label{eq:11c}
P= \oint_{\Sigma}{{\rm d}\Sigma}\:({\bf n} \cdot {\cal S})
=\frac{e^2[\dot{{\bf v}}^{2}]_{ret}}{2 c^3}\int_{\rm o}^{\pi} {\rm d}\theta\: \sin^3\theta
=\frac{2e^{2}}{3c^{3}} [\dot{\bf v}^{2}]_{ret}.
\end{equation}
This Poynting flux through the spherical surface is independent of its radius $r$, and one infers that this radiated power crossing the 
surface at time say, $t_{\rm o}$, must equal the mechanical energy loss rate ($\propto \dot{{\bf v}}^2$) of the charge at the retarded time 
$t_{\rm o}-r/c$, when it had an acceleration $\dot{{\bf v}}$. This is Larmor's famous result for an accelerated charge that the radiative  
power loss at any time is proportional to the square of its acceleration at that instant; the power loss occurring  
presumably out of the kinetic energy of the charge. 

It is usually assumed that the acceleration fields, which fall with distance as $1/r$, solely represent the radiation from a charge, 
since the contribution of the velocity fields ($\propto 1/r^{2}$) appears to be negligible for a large enough value of $r$. 
Of course that would not be the case if the velocity were also changing monotonically with time, as for example in the case of a uniform 
acceleration where the {\em retarded value} of the velocity will be having a term ${\bf v} \propto -\dot{\bf v} r/c$ 
and in order to calculate the field at any event (space and time location), for a larger $r$, 
we will be going proportionally further back in retarded time ($t_{\rm o}-r/c$) and consequently the term 
${\bf v}/cr^2$ in the velocity fields might become comparable to the acceleration field term $\dot{\bf v}/c^2r$ in Eq.~(\ref{eq:1a})). 
In fact this makes it imperative in Larmor's derivation that the motion considered is in a sense repetitive (though not necessarily 
in regular cycles), where the velocity will not be increasing monotonically with acceleration and thence the acceleration 
too would not be a constant and will be having a finite time-derivative. 
\section{A mathematical subtlety missed in the derivation of Larmor's formula}
In the text-book derivation of Larmor's formula a mathematical subtlety seems to have hitherto been overlooked while applying 
Poynting's theorem which, strictly speaking, relates the outgoing radiation flux from 
a closed surface to the rate of energy decrease within the enclosed volume, all calculated only {\em for the same instant of time}. 
While deriving Larmor's formula (Eq.~(\ref{eq:11c})), one equated the Poynting flux through a surface at a time $t_{\rm o}$ to the 
radiation loss of the charge {\em at the retarded time} $t_{\rm o}-r/c$. It is true that the electromagnetic fields at $r$ at time $t_{\rm o}$ 
are determined by the charge position and motion at the retarded time $t_{\rm o}-r/c$, and one might expect a similar causal relation 
for the radiated power. However, the relation between the Poynting flux through a sphere to the rate of the energy loss in the enclosed volume, 
is given by Poynting's theorem which is valid strictly when all quantities are evaluated for {\em the same instant of time}. 
Poynting's theorem does not state any relation between the radiation flux at $t_{\rm o}$ and the energy loss rate of the charge at 
$t_{\rm o}-r/c$. We may make $r$ as small as we want but $t_{\rm o}>t_{\rm o}-r/c$ always. 

Let a charge moving with a constant acceleration $\dot{\bf v} $ be instantly stationary (i.e., ${\bf v}=0$)  
at a time $t=0$, then we can write for the electromagnetic fields,
\begin{equation}
\label{eq:1a1}
{\bf E}=\left[\frac{e\:{\bf n}}{r^{2}} + \frac{e\:{\bf n}\times\{{\bf n}\times \dot{\bf v}\}}{rc^2}\right]_{t=0}
\end{equation}
Equation ~(\ref{eq:1a1}) gives for any time $t=\tau$ the 
electromagnetic field on a spherical surface $\Sigma$ of radius $r=c \tau$ centred on the charge position at $t=0$. 
Then Larmor's formula says that the Poynting flux through the spherical surface at time $t=\tau$, 
\begin{equation}
\label{eq:1a1a}
P=\frac{2e^{2}}{3c^{3}} \dot{\bf v}^{2}
\end{equation}
is the energy loss rate of the charge at the retarded time $t=0$. However that does not seem consistent with the law of energy 
conservation since the charge was instantly stationary and had a nil kinetic energy at $t=0$ (we do not entertain the possibility that 
the radiation losses may be out of the proper mass of the charge, though such a thing has been considered in the literature \cite{8}).

Poynting's theorem relates the mechanical work done on the charge by the 
electromagnetic fields (which in this case comprise its self-fields as there are no other fields or charges within the system 
being considered here) to the Poynting flux through a surface enclosing the charge, both quantities evaluated for the same time, 
say $t_{\rm o}$. Now the fields at $t_{\rm o}$ are determined by the motion of the charge at the retarded time $t_{\rm o}-r/c$. 
Thus, in order to calculate the power loss by an instantly stationary charge at $t=0$ 
one must consider the Poynting flux through a surface surrounding the charge at $t=0$ itself, even though the fields  
at the surface are determined from the time retarded positions of the charge when it presumably was not yet 
stationary, say, at $t=-r/c$. As we shall show later the velocity and acceleration fields at the surface at time $t=0$,  
arising from the retarded position of the charge at $t=-r/c$ neatly cancel each other, resulting in a nil Poynting flux through 
the surface, consistent with the charge possessing a nil kinetic energy at the time $t=0$.

Further, in the derivation of  Larmor's formula one calculated the radiated power from Poynting flux due to acceleration fields
under the assumption that these contribute exclusively to the radiation and ignored the velocity fields which may also have a transverse 
component and thus a finite Poynting flux. Actually as the charge undergoes an acceleration, the energy in the self-fields in its 
neighbourhood must be "co-moving" with the changing velocity of the charge (after all the self-fields cannot be lagging behind the charge) 
and there would be a Poynting flux due to that. Therefore not all of the Poynting flux may constitute radiation; 
some of it is just field energy dragged along the particle as it moves. The radiated energy is
the part that propagates off to infinity and is detached from the charge \cite{25}, i.e., it no longer remains 
a part of the self-fields of the charge. 

It is well known that the self-field energy of a charge moving with a uniform velocity is different
for different values of the velocity (see e.g.~\cite{15}). After all when a charge is accelerated, depending upon the 
change in velocity, its self-field energy must change too. But that change in self-field energy cannot come from the velocity fields 
alone (the first term on the right hand side of Eq.~(\ref{eq:1a})) which contains no information about the change that might take  
place in the velocity of the charge. Therefore 
the acceleration fields, to some extent at least, must represent the changes taking place in the energy in self-fields which are  
attached to the charge. On the other hand, in the standard picture of ``radiation'', acceleration fields are considered, exclusively and wholly, 
to represent power that is ``lost'' by the charge irreversibly as radiation and that of course is the genesis of Larmor's 
formula of radiative losses. Thus we see that there is something amiss in the standard picture which does not take into account 
whatsoever contribution of the acceleration fields towards the changing self-field energy of the accelerating charge. Actually 
radiation will be that part of the Poynting flux which is over and above the flux value arising from the ``present''  velocity of the charge. 

We can find out the Poynting flux from the self-fields being ``dragged'' along the charge due to its "present" velocity 
${\bf v}_{\rm o}$, by a comparison with the Poynting flux of a charge moving with a constant (non-relativistic) velocity,  
${\bf v}_{\rm o}$ (i.e., $\dot{\bf v}=0$), for which we get the transverse component of the 
electric field from Eq. (\ref{eq:1a}) as,
\begin{equation}
\label{eq:1q}
{\bf n}\times{\bf E}=\frac{-e{\bf n}\times {\bf v}_{\rm o}}{r^{2}c}\;.
\end{equation}
Accordingly we get Poynting flux arising from the velocity fields of a charge moving with a velocity ${\bf v}_{\rm o}$ as,
\begin{equation}
\label{eq:1r}
\frac{e^2c}{2}\int_{\rm o}^{\pi} {\rm d}\theta\: \sin^3\theta \:r^2
 \left[\frac{{\bf v}_{\rm o}}{r^{2}c}\right]^{2}=\frac{2e^2{\bf v}_{\rm o}^2}{3r^{2}c}\;,
\end{equation}
which is true for all values of $r$. 

The Poynting flux in Eq.~(\ref{eq:1r}) at time $t$ due to the "present" velocity of the accelerating charge 
can be related to the erstwhile called radiated power (Eq.~(\ref{eq:1a1a})) by substitution of ${\bf v}_0 =\dot{\bf v}r/c$ 
(for a constant acceleration as we assumed earlier) to get,  
\begin{equation}
\label{eq:1r1}
\frac{2e^2{\bf v}_{\rm o}^2}{3r^{2}c}=\frac{2e^2\dot{\bf v}^2r^{2}}{3r^{2}c^3}=\frac{2e^2\dot{\bf v}^2}{3c^3} \;.
\end{equation}
Thus we see that in the case of a constant acceleration, the hitherto termed radiation losses (\`{a} la Larmor's formula) 
are nothing but the Poynting flow due to the movement of self-fields along with the
charge, because to its "present" velocity ${\bf v}_{\rm o}$.    

\section{Poynting flux in the neighbourhood of a moving point charge}
Using the vector identity ${\bf v}={\bf n}({\bf v}.{\bf n}) - {\bf n}\times\{{\bf n}\times{\bf v}\}$, 
we can decompose the electric field in Eq.~(\ref{eq:1a}), 
in terms of the radial (along {\bf n}) and transverse components as \cite{17,18},
\begin{eqnarray}
\label{eq:11a}
{\bf E}=\left[\frac{e{\bf n}}{\gamma ^2 r^2(1-{\bf n}\cdot{\bf v}/c)^2}\right.
+\left.\frac{e{\bf n}\times\{({\bf n}-{\bf v}/c)\times({\bf v}+\gamma^2 \dot{\bf v}r/c)\}}
{\gamma^2r^2c\:(1-{\bf n}\cdot{\bf v}/c)^3}\right]_{ret},
\end{eqnarray}
the quantities on the right hand side evaluated 
at the retarded time. 

This is a general expression for the electric field of a charge, 
with the radial and transverse terms fully separated.   
The second term on the right hand side 
includes transverse terms both from the velocity and acceleration
fields together, that should be responsible for the net Poynting flux through a surface surrounding the charge. 

We assume that the motion of the charged particle is non-relativistic and it varies slowly so that during the light-travel time
across the particle, any change in its velocity, acceleration or other higher time derivatives
is relatively small. This is equivalent to the conditions
that $|{\bf v}|/c \,\ll\,1, \;|\dot{{\bf v}}|r_{\rm o}/c \,\ll\,c, \;|\ddot{{\bf v}}|r_{\rm o}/c \,\ll 
\,|\dot{{\bf v}}|,$ etc. Therefore we keep only 
linear terms ${{\bf v}}$, $\dot{{\bf v}}$ etc., in our formulation. 
Then from Eq.~(\ref{eq:11a}) we can write,
\begin{equation}
\label{eq:12h}
{\bf n}\times {\bf E}=-e\left[\frac{{\bf n}\times ({\bf v}+\dot{\bf v}r/c)}{r^{2}c}\right]_{ret}\;.
\end{equation}
Then using Eq.~(\ref{eq:12h}) for the transverse electric field, 
one gets the electromagnetic power passing through the surface $\Sigma$ as, 
\begin{equation}
\label{eq:1n}
P= \oint_{\Sigma}{{\rm d}\Sigma}\:({\bf n} \cdot {\cal S})
= \frac{e^2c}{2}\int_{\rm o}^{\pi} {\rm d}\theta\: \sin^3\theta \:r^2
\left[\frac{({\bf v}+\dot{\bf v}r/c)}{r^{2}c}\right]_{ret}^{2}\;.
\end{equation}
or 
\begin{equation}
\label{eq:1o}
P= \frac{2e^2}{3c}\left[\frac{({\bf v}+\dot{\bf v}r/c)^2}{r^{2}}\right]_{ret}\;,
\end{equation}

Here the Poynting flux through a surface at a time $t_{\rm o}$ is written in terms of the  
charge motion {\em at the retarded time} $t_{\rm o}-r/c$, i.e., ${\bf v}$ and $\dot{\bf v}$ in Eq.~(\ref{eq:1o}) 
are values at the retarded time. 
Also if we ignore velocity ${\bf v}$ in Eq.~(\ref{eq:1o}), we get usual Larmor's formula (c.f. Eq.~(\ref{eq:11c})).

As we discussed above, to correctly apply the Poynting theorem, we must express Eq.~(\ref{eq:1o}) 
in terms of the charge motion at $t_{\rm o}$.
For this we can make a Taylor series expansion of the velocity and acceleration at the retarded time $t_{\rm o}-r/c$ 
in terms of the charge motion at present time $t_{\rm o}$ as, 
\begin{equation}
\label{eq:1d}
{{\bf v}}={{\bf v}}_{\rm o}-\frac{\dot{{\bf v}}_{\rm o}r}{c}+\frac{\ddot{{\bf v}}_{\rm o}r^{2}}{2c^{2}}+\cdots,
\end{equation}
\begin{equation}
\label{eq:1f}
\dot{{\bf v}}=\dot{{\bf v}}_{\rm o}-\frac{\ddot{{\bf v}}_{\rm o}r}{c}+\cdots.
\end{equation}
all quantities on the right hand side evaluated at the present time $t_{\rm o}$. 

Substituting for ${\bf v}$ and $\dot{\bf v}$ from Eqs.~(\ref{eq:1d}) and (\ref{eq:1f}) into 
Eq.~(\ref{eq:1o}), we get,
\begin{equation}
\label{eq:1o1}
P= \frac{2e^2({\bf v}_{\rm o}-\ddot{\bf v}_{\rm o}r^2/2c^2)^2}{3r^{2}c}\;,
\end{equation}
where we notice that the acceleration dependent term does not appear in the power formula as it gets cancelled.
We should point out here that the spherical surface $\Sigma$ of radius $r$ considered here, is centred around the retarded 
position of the charge, which is somewhat off from the present position of 
the charge due to its motion. However, the charge and its neighbourhood is still enclosed well within the 
surface $\Sigma$ (for $|{\bf v}|/c \,\ll\,1$) and Poynting's theorem is as much applicable to the radiation flux through this surface 
as to a surface centred on the charge.

Now dropping terms of order $r^2$ or its higher powers, that will become zero as $r\rightarrow 0$, 
we get, 
\begin{equation}
\label{eq:1o2}
P= \frac{2e^2{\bf v}_{\rm o}^2}{3r^{2}c}-\frac{2e^2{\bf v}_{\rm o}\cdot\ddot{\bf v}_{\rm o}}{3c^{3}}\;.
\end{equation}
This is the Poynting flux in the neighbourhood of a point charge, instead of Larmor's formula as given by  Eq. (\ref{eq:11c}). 
The first term on the right hand side is the self-Coulomb field energy of the charge moving with a ``present'' velocity ${\bf v}_{\rm o}$ 
( as per Eq.~(\ref{eq:1r})) while second term is the same as the power loss derived in the literature earlier because of drag against 
self-force of an accelerated charged sphere \cite{16,7,24,20}.
\section{Radiation losses from a charge in arbitrary motion}
As we mentioned earlier, in order to relate the Poynting's flux to the power loss by the charge, we should distinguish between the 
Poynting flux due to the self-fields carried along the charge due to its ``present'' velocity and the remainder not represented in the motion 
of the charge and is thus detached from it and represents irretrievable radiative losses from the charge. The first term on the right hand side 
in Eq.~(\ref{eq:1o2}) is the Poynting flow due to the movement of self-fields along with the charge, 
because to its ``present'' velocity ${\bf v}_{\rm o}$, as the Poynting flux is the same as in Eq.~(\ref{eq:1r}) for all values of $r$. 
Of course both expressions diverge as $r\rightarrow 0$, but that is unavoidable as it is due to the ``point'' nature of the 
considered charge distribution, 
and the same divergence to infinity  for $r\rightarrow 0$ is present even in the Coulomb field energy of a stationary charge. 
Thus in Eq.~(\ref{eq:1o2}) it is only the remainder, distance-independent second term, not contained in the velocity fields 
of the charge, that could therefore be considered to be detached from the charge and hence radiated away.

In Eq.~(\ref{eq:1r1}) the charge had a zero initial velocity. For a non-zero initial velocity ${\bf v}$ the Poynting flux in 
Eq.~(\ref{eq:1o}) is again due to the "present" velocity, ${\bf v}_0 ={\bf v}+\dot{\bf v}r/c$, provided the acceleration $\dot{\bf v}$ is a constant. 
Here there is no excess flux 
(than what required for its present velocity) that could be called as radiative loss. In fact in such a case even if we let $r$ become 
very large, the Poynting flux will still represent the large velocity of the charge unless there is a change in acceleration, and it is 
only in the latter case that the Poynting flux at large $r$ will not match the value required for the actual motion 
(with a changing acceleration) of the charge at 
that time. This remains true even if the charge achieves a relativistic motion. For example, in the case of a uniform acceleration, where 
the charge may acquire relativistic velocity for a large enough time (we assume a one-dimensional motion with 
${\bf v}\parallel \dot{\bf v}$), then instead of Eq.~(\ref{eq:1o}) we get for the Poynting flux \cite{18}, 
\begin{equation}
\label{eq:1r2}
P= \frac{2e^2(\gamma {\bf v}+ \gamma^{3} \dot{\bf v}{r}/{c})^2}{3r^{2}c}=\frac{2e^2{(\gamma {\bf v})_{\rm o}^2}}{3r^2c}\;,
\end{equation}
which is nothing but a relativistic generalization of Eq.~(\ref{eq:1r}),  
since $\gamma {\bf v}+ \gamma^{3} \dot{\bf v}{r}/{c}=(\gamma {\bf v})_{\rm o}$ 
for a uniform acceleration.

From detailed analytical calculations it has been shown \cite{17,18} that for a uniformly accelerated charge, the total 
energy in the fields (i.e., including both the velocity and acceleration field terms from Eq.~(\ref{eq:1a})) at any 
time is just equal to the self-energy of a charge moving uniformly with a velocity equal to the 
instantaneous ``present'' velocity of the accelerated charge (even though the detailed field configurations may differ in the two cases). 
In fact as we showed above, for a uniformly accelerated charge (with $\ddot{\bf v} = 0$), all the Poynting flux in the 
acceleration fields goes towards making the change in the self-field energy of the charge as its velocity changes due to the acceleration. 

Thus only when a rate of change of acceleration is present that we get excess Poynting flux than that needed for the present velocity  
of the charge and this excess electromagnetic power constitutes the radiative loss by the charge.
The standard Larmor's expression, in general, comprises the Poynting flux even of the self-field corresponding to the ``present'' 
motion of the accelerating charge, along with the radiative losses, if any.

According to the Poynting's theorem the rate of change of the mechanical energy $({\cal E_{\rm me}})$ of the 
charges plus the electromagnetic field energy $({\cal E_{\rm em}})$ enclosed 
within a volume could be equated to the negative of the net Poynting flow through a surface surrounding the volume, 
with {\em all quantities evaluated at the same time} $t_o$. 
\begin{equation}
\label{eq:21a}
\frac{d{\cal E_{\rm me}}}{{\rm d}t}+\frac{d{\cal E_{\rm em}}}{{\rm d}t}=-\int_{\Sigma}{{\rm d}\Sigma}\:({\bf n} \cdot {\cal S}) \;.
\end{equation}

Now in case of a "point" charge, the volume integral of electromagnetic energy,  $({\cal E_{\rm em}})$, diverges. However, we can instead use a 
comparison 
with a uniformly moving charge with velocity equal to the "present" velocity of the accelerated charge. It has been explicitly shown \cite{17,18} 
that for a uniformly accelerated charge, the field energy as well as Poynting flux are equal to that of a uniformly 
moving charge with velocity equal to the "present" velocity of the uniformly accelerated charge, and there is no other flux that could be 
called radiation loss. Thus it is only the excess Poynting flux due to rate of change of acceleration that represents radiative power loss. 
Therefore, from a comparison with Eq.~(\ref{eq:1r}), we infer that it is only the second term in Eq.~(\ref{eq:1o2}), viz.,
\begin{equation}
\label{eq:1s}
P= \frac{-2e^2\ddot{\bf v}_{\rm o}\cdot{\bf v}_{\rm o}}{3c^{3}}\;,
\end{equation}
which is the excess outgoing power from the charge and therefore represents instantaneous radiative losses. 
Larmor's formula does not separate out, from the radiated power, the contribution of acceleration fields (the 2nd term on the right hand side of 
Eq.~(\ref{eq:1a})) to the Poynting flux that is due to the changing energy in velocity fields. Of course in vast majority of cases where the 
energy in velocity fields does not keep on increasing indefinitely, Larmor's formula yields a time-averaged value of power loss 
undergone by the radiating charge.

The expression for power loss (Eq.~(\ref{eq:1s})) is the same as hitherto obtained in the literature from 
a detailed derivation of the self-force of an accelerated charge sphere of a vanishingly small radius \cite{16,7,20}.
But here we have derived the power loss due to radiation reaction from the Poynting flow for a ``point source'' and which we showed 
to be different from the familiar Larmor's formula. 
This should now also obviate the need for the internal-energy term, introduced on an ad hoc 
basis \cite{7,41,56}, with a desire to make the power loss due to radiation reaction comply with the instantaneous 
radiative losses, hitherto thought to be given by Larmor's formula.

We can generalize the formula for radiative losses to a relativistic 
case, by using the condition of relativistic covariance (see e.g.,~\cite{2,3}),
\begin{equation}
\label{eq:10}
P=-\frac{2e^{2}\gamma_{\rm o} ^{4}}{3c^{3}}\left[\ddot{\bf v}_{\rm o}\cdot{\bf v}_{\rm o}+
3\gamma_{\rm o} ^{2}\frac{(\dot{\bf v}_{\rm o}\cdot{\bf v}_{\rm o})^{2}}{c^{2}}\right]\:.
\end{equation}
Equation~(\ref{eq:10}) should be contrasted with Li\'{e}nard's formula for radiative power losses from a charge moving with an arbitrary 
velocity, obtained by a relativistic transformation of Larmor's formula \cite{1,25,75}, 
\begin{equation}
\label{eq:11}
P=\frac{2e^{2}\gamma ^{6}}{3c^{3}}\left[\dot{\bf v}^{2}-
\frac{({\bf v}\times\dot{\bf v})^{2}}{c^{2}}\right]
=\frac{2e^{2}\gamma ^{4}}{3c^{3}}\left[\dot{\bf v}^{2}+
\frac{\gamma ^{2}({\bf v}\cdot\dot{\bf v})^{2}}{c^{2}}\right].
\end{equation}
Equations (\ref{eq:10}) and  (\ref{eq:11}) apparently look very different. Later we shall explore 
the difference that the two formulas make in their applicability.
\section{The absence of radiation from an instantly stationary point charge}
From Eqs.~(\ref{eq:1s}) or (\ref{eq:10}), $P=0$ if ${{\bf v}}_{\rm o}=0$, which means there is no radiation from an instantaneously stationary point 
charge. Actually it is evident from Eqs.~(\ref{eq:1o2}) itself that 
there is a nil Poynting flux through a surface $\Sigma$ around the charge for ${{\bf v}}_{\rm o}=0$. 
In fact, there is no Poynting flux anywhere in the entire vicinity of the instantaneously stationary charge. 
From this one readily infers an answer to the question whether 
for such a charge there are any energetic or radiative losses at that moment, then the answer is an emphatic NO. 
The expected radiation flux term from Larmor's formula, proportional to the square of the acceleration and independent of the 
radius of the sphere ($=2e^2\dot{{\bf v}}_{\rm o}^2/3c^3$), is certainly not present in the case of an instantly stationary charge. 
In all neighbourhood of the charge in its instantaneous rest-frame, the transverse terms of the time-retarded velocity fields 
cancel the acceleration fields which were responsible in Larmor's formula for a distance-independent radiation  
power through a surface surrounding the charge. From the absence of any Poynting flux in its neighbourhood it is clear that 
no net energy loss is taking place by the charge at that instant.

It is interesting to note that in the case of a uniformly accelerated charge, the acceleration fields get cancelled completely by the 
transverse term of the velocity fields not merely in the neighbourhood but {\em at all distances}, in its instantaneous rest-frame.  
For a uniformly accelerated charge, even when the motion might have been relativistic at the retarded time, for all values of $r$ 
the time-retarded value of the expression $\gamma {\bf v}+ \gamma^{3} \dot{\bf v}r/c$ represents the ``present'' 
velocity, $(\gamma {\bf v})_{\rm o}$, which becomes zero in the instantaneous rest-frame of the charge. 
Actually for a larger $r$, we need to go further back in time to get the time-retarded value of velocity which is 
directly proportional to $t=r/c$ for a uniform acceleration. This results in,
\begin{equation}
\label{eq:11b}
\gamma {\bf v}+ \gamma^{3} \dot{\bf v}\frac{r}{c}=(\gamma {\bf v})_{\rm o}=0\;,
\end{equation}
with all transverse field in Eq.~(\ref{eq:11a}) getting cancelled for all $r$, implying zero Poynting flux and hence nil radiation 
in the instantaneous rest-frame. This argument was used to show the absence of radiation for a charge supported in 
a gravitational field \cite{17,18}, in conformity with the strong principle of equivalence. 
Incidentally Pauli \cite{33} first brought it to notice that magnetic field is zero throughout 
in the instantaneous rest-frame of a uniformly accelerated charge, indicating the absence of radiation in this case. 

Thus one arrives at the conclusion that there is a nil radiative loss rate from a charge that is instantaneously stationary, 
in conformity with the requirement that in the instantaneous rest-frame of the charge, because of a nil kinetic energy and 
a nil rate of work being done on the charge by the external agency, responsible for the acceleration of the charge,   
nothing could have provided necessary power that can go into the radiation. 

Here, however, a possible objection could be raised. After all for any motion of the charge, at any time one can find an inertial frame in which the 
charge is momentarily at rest, and thus not radiating according to the above arguments. Does it mean that the charge does not radiate at all? 

Actually one could look at this intriguing problem from two different aspects. The first one is the "power loss" by the accelerated "point" charge, 
$|{\rm d}(1/2 mv^2)/{\rm d}t|=|m v {\rm d}v/{\rm d}t|$, which will be zero when $v=0$, even if only for an instant in the instantaneous rest frame. 
But it will be finite in a frame where $v$ is finite. The second 
one is from an observer's point of view. After all one may rightly object that the energy in electromagnetic fields may be transformed but 
cannot be made totally zero by changing to any frame if it is finite as measured 
at least in one other frame, of say, a distant observer.  To understand this let us say one observer (in the instantaneous rest frame) instantly 
measures a zero Poynting flux through a spherical surface. Now another observer (in a moving frame) will see these as not simultaneous measurements, 
but made at different times at different parts of the spherical surface, and in his own set of simultaneous measurements (that is in second 
observer's time) could still see a finite flux passing through the surface. Thus field energy-momentum, which is volume integrals over extended 
regions of space, even if it turns out instantly to be zero in one frame, need not be necessarily be zero in other frames too. 
\section{Applicability of the old versus new formulas}
Recently it has been shown \cite{68} that Larmor's formula is compatible with the power loss from radiation reaction and that 
the apparent discrepancy in the two formulations arises only because the two are inadvertently 
expressed in two different time-systems. If the motion of the charge is expressed in terms of the values of its parameters 
(velocity, acceleration and higher time-derivatives, if any) at the present time, 
then we arrive at the formula for radiation losses due to radiation reaction. On the other hand, 
expressing the charge motion in terms of the parameter values at the retarded time gives us the familiar Larmor's radiation 
formula. In particular, it was explicitly shown \cite{68} that for a charge moving in a circle (e.g., in a cyclotron or a synchrotron case), 
the two formulas yield the same power loss rate at every instant. 
The instantaneous power rate is identical in the case of {\em a circular motion} can be directly seen from Eqs. (\ref{eq:10}) and (\ref{eq:11}) 
as well, because $\dot{\bf v}\cdot{\bf v}=0$ also implies $\ddot{\bf v}\cdot{\bf v}+\dot{\bf v}^{2}=0$, and 
then the two expressions for power losses (Eqs. (\ref{eq:10}) and  (\ref{eq:11})) yield equal values.

Of course this compatibility does not mean that the two formulas, in general, give identical results. 
In order to understand the difference in the two formulas when applied to a charge with an arbitrary motion, 
let us first consider a harmonically oscillating charge (like in a radio antenna).  
\begin{equation}
\label{eq:10.1}
{\bf x}={\bf x}_0 \cos(\omega t+\phi)\;.
\end{equation}
Then
\begin{equation}
\label{eq:10a}
{\bf v}=\dot{{\bf x}}=-\omega {\bf x}_0 \sin(\omega t+\phi)\;,
\end{equation}
\begin{equation}
\label{eq:10b}
\dot{{\bf v}}=\ddot{{\bf x}}=-\omega^2 {\bf x}_0 \cos(\omega t+\phi)=-\omega^2 {\bf x}\;,
\end{equation}
\begin{equation}
\label{eq:10c}
\ddot{{\bf v}}=\omega^3 {\bf x}_0 \sin(\omega t+\phi)=-\omega^2 {\bf v}\;.
\end{equation}

Then Larmor's formula yields radiative power $\propto \dot{\bf v}^2=\omega^4 {\bf x}^2_0 \cos^2(\omega t+\phi)$ while the power loss from 
the radiation reaction turns out $\propto -\ddot{\bf v}\cdot{\bf v}= \omega^4 {\bf x}^2_0 \sin^2(\omega t+\phi)$. Though the two  
expressions yield equal radiated energy when integrated or averaged over a complete cycle, the instantaneous rates are quite different, 
in fact the two rates of power loss will be out of phase with each other. When the instantaneous 
power loss as calculated from the radiation reaction will be 
maximum, the radiation loss estimated from Larmor's formula will be minimum and vice versa. For instance, when $\omega t+\phi=0$ and $v=0$ 
from Eq. (\ref{eq:10a}), the radiation reaction equation gives zero power loss, Larmor's formula yields maximum power loss rate. 
Thus the two formulas may give the same result 
for the power loss in a time averaged sense, however the strictly instantaneous rates could be very different. 
As any actual motion of the charge could be Fourier analysed, the above statement would be true for individual Fourier components 
though the detailed time behaviour of the combined effect could be quite different. However, in spite of this difference, 
the time averaged power losses would be similar. Also it implies that the 
power spectrum, which gives average power in the cycle for each frequency component, would be the same.
Of course there will be cases where a Fourier analysis is not possible, for 
instance, in the case of a uniformly accelerated charge. In such cases the two formulas could yield conflicting answers. 

\section{Radiation reaction in a synchrotron source}
It might seem that the difference between radiative losses given by Larmor's formula or those inferred from the 
radiation reaction formula might be more of an academic 
interest or at most making a difference only in some very special cases like that of a uniformly accelerated charge. 
First thing, uniformly accelerated charges, are not a rare breed. All charges stationary in a 
gravitation field of a star or other celestial bodies, including that of Earth, belong to this category. Secondly there 
could be other cases where a reasoning based on Larmor's formula (or its relativistic generalization Li\'{e}nard's result) 
could lead to potentially wrong conclusions. An example is the calculation of dynamics of relativistic 
charges radiating in a synchrotron source (be it an astrophysical or a laboratory phenomenon).

A charge in a constant uniform magnetic field moves in a helical path with a velocity component 
parallel to the magnetic field ${\bf v}_{\parallel}={\bf v} \cos \theta$, that remains unaffected by the magnetic field. 
Here $\theta$ is the pitch angle, i.e., angle of the velocity vector of the charge with respect to the magnetic field vector. 
Now all the radiated power from a highly relativistic charge in a synchrotron case, as calculated from Larmor's formula 
(or rather from Li\'{e}nard's formula), 
is within a narrow cone of angle $1/\gamma$ around the instantaneous direction of motion \cite{1,26,27}. 
Therefore any radiation reaction on the charge would be in a direction opposite to its instantaneous velocity vector \cite{23}. 
This means that the direction of motion of the charge will not be affected, implying no change in 
the pitch angle of the charge. The subsequent formulation depends on these arguments and the dynamics as well as the life times 
of the synchrotron electrons are accordingly calculated. The formulas have appeared in various review articles and 
text-books \cite{26,72,73}, and are widely in use for the last 50 years.

However, there is something amiss in the above arguments and it turns out that this picture is not consistent with the 
special theory of relativity. In a synchrotron case, component 
${\bf v}_\parallel$ of a charge remains constant, while magnitude of ${\bf v}_\perp$ decreases continuously and as a consequence the 
pitch angle of the radiating charge in general changes \cite{31,31a}.
This can be seen in the gyro-centre (G.C.) frame, which moves with a velocity ${\bf v}_{\parallel}$ 
with respect to the lab-frame ${\cal S}$ and in which 
the charge therefore has only a circular motion in a plane perpendicular to the magnetic field (with a pitch angle $\theta=90^\circ$). 
In the G.C. frame, due to radiative losses, there will be a decrease in the velocity, which is solely  
in a plane perpendicular to the magnetic field.

Now $v_\parallel$ is the constant relative velocity between two reference frames, therefore even in the lab frame, the parallel component 
of velocity of the charge should remain unchanged. However, magnitude 
of the perpendicular component of velocity is continuously decreasing because of radiative losses, therefore the pitch angle of 
the charge given by, $\tan\theta = {\bf v}_\perp/{\bf v}_\parallel$, should decrease continuously with time and 
the velocity vector of the charge should increasingly align with the magnetic field vector. 
While from Larmor's formulation it has been inferred that the pitch angle of the radiating charge 
does not change, from special relativistic arguments we conclude that the pitch angle of the radiating charge 
would continuously decrease, with the velocity vector of the charge gradually getting aligned with the magnetic field 
direction \cite{31a}. 
A careful examination of the effect of the radiation reaction in synchrotron case on the dynamics of the charged particle,  
viz. the pitch angle of the radiating charge decreases, which appears contrary to the conventional wisdom, but is in complete agreement with 
conclusions based on special relativistic arguments \cite{31,31a}. 

It is quite intriguing that in spite of the instantaneous power loss rate being the same in the two cases, the effect on the charge motion 
is quite different in each case. In particular Larmor's formula does not yield results compatible with special relativity. 
There is actually an inherent inconsistency in estimating radiation reaction on a charge from Larmor's formula.  
The radiation pattern of an accelerated charge has a $\sin^2\phi$ dependence 
about the direction of acceleration \cite{1,2,25}. Due to this azimuthal symmetry 
the net momentum carried by the radiation is nil. Therefore the charge too cannot be losing momentum, even though it is undergoing radiative losses. 
Thus we have a paradox of a radiating charge losing its kinetic energy but without a corresponding change in its linear momentum. 
Such a paradox does not appear in the radiation reaction formulation, which alone seems to beget results consistent with the 
special relativity.
\section{Conclusions}
From the electromagnetic fields in the neighbourhood of a ``point charge'' in arbitrary motion, we determined   
the Poynting flux across a spherical surface of vanishingly small radius surrounding the charge. From that we  
showed that the radiative power loss turns out to be proportional to the scalar product of the instantaneous velocity and the 
first time-derivative of the acceleration of the charge. 
The discordance of these results with the familiar Larmor's formula was traced to the facts that firstly, in the text-book derivation of 
Larmor's formula, one did not properly consider the contribution of the velocity-dependent self-fields to the Poynting flux. Secondly,  
one equated the Poynting flux through a surface at a later time to the radiation loss by the charge 
at the retarded time, which is not what Poynting's theorem states. 
A consistent picture was shown to emerge only when a mathematical correct procedure is followed.
We showed that Larmor's formula in general, gives the radiative power loss only in a time-averaged sense and 
does not always yield an {\em instantaneous} value of radiative loss from a point charge. 
In particular, even for an instantly stationary accelerated charge, 
Larmor's formula predicts a finite rate of radiative losses proportional to 
the square of acceleration of the charge, in violation of energy conservation as the charge has no kinetic energy that could be 
lost into radiated power. However, a proper examination of the electromagnetic fields in the neighbourhood of the point charge and 
the Poynting flux across a surface surrounding the point charge shows the absence of any such radiation flux in the instantaneous 
rest frame, where the contribution of acceleration fields is cancelled by the velocity fields.  
Further, we showed that in the case of synchrotron radiation, Larmor's formula leads to a potentially wrong conclusion  
about the constancy of the pitch angle of a radiating charge, a result  
inconsistent with the special relativity, and that only the radiation reaction formula yields a correct result 
about the decrease in pitch angle due to radiation losses, with the velocity vector getting gradually aligned 
with the magnetic field direction.
{}
\end{document}